\documentclass[preprint2]{aastex61}

\received{}
\revised{}
\accepted{}
\submitjournal{ApJ}

\shorttitle{Optical, near-IR  and $X$-ray observations of SN 2015J and its host galaxy}
\shortauthors{Nucita et al.}

\usepackage[utf8]{inputenc}
\usepackage{graphicx,epsfig}  
\usepackage{amsmath,amssymb}
\usepackage{longtable}
\usepackage{times}
\usepackage{textcomp}
\usepackage{natbib}
\usepackage{txfonts}
\usepackage{fixltx2e}

\def\ut#1{\mathop{\vtop{\ialign{##\crcr
     $\hfil\displaystyle{#1}\hfil$\crcr\noalign
     {\kern1pt\nointerlineskip}\hbox{$\hfil\sim\hfil$}\crcr
     \noalign{\kern1pt}}}}}

\def\undersymbol#1#2{\mathop{\vtop{\ialign{##\crcr
     $\hfil\displaystyle{#2}\hfil$\crcr\noalign
     {\kern1pt\nointerlineskip}\hbox{$\hfil#1\hfil$}\crcr
     \noalign{\kern1pt}}}}}
\def\arcsec{^{\prime\prime}}

\def\degr{^\circ}
\def\hour{^{\rm h}}
\def\minute{^{\rm m}}
\def\second{^{\rm s}}

\begin{document}

\title{Optical, near-IR  and $X$-ray observations of SN 2015J and its host galaxy\footnote{
Based on observations obtained with XMM-Newton, an ESA science mission with instruments and contributions directly funded by ESA Member States and NASA, 
with ESO Telescopes at the La Silla-Paranal Observatory under programme ID 298.D-5016(A), and with the 6.5 meter Magellan Telescopes located at Las Campanas Observatory, Chile.
We also acknowledge the use of public data from the Swift data archive.}}

\correspondingauthor{A. A. Nucita}
\email{nucita@le.infn.it}

\author{A. A. Nucita}
\affil{Department of Mathematics and Physics {\it ``E. De Giorgi''}, University of Salento, Via per Arnesano, CP 193, I-73100,
  Lecce, Italy}
\affil{INFN, Sez. di Lecce, Via per Arnesano, CP 193, I-73100, Lecce, Italy}

\author{F. De Paolis}
\affiliation{Department of Mathematics and Physics {\it ``E. De Giorgi''}, University of Salento, Via per Arnesano, CP 193, I-73100,
  Lecce, Italy}
\affiliation{INFN, Sez. di Lecce, Via per Arnesano, CP 193, I-73100, Lecce, Italy}

\author{R. Saxton}
\affiliation{European Space Astronomy Centre, SRE-O, P.O.~Box 78, 28691, Villanueva de la Ca\~nada (Madrid), Spain}

\author{V. Testa}
\affiliation{INAF, Osservatorio Astronomico di Roma, via Frascati 33, I-00078 Monte Porzio Catone, Italy}

\author{F. Strafella}
\affiliation{Department of Mathematics and Physics {\it ``E. De Giorgi''}, University of Salento, Via per Arnesano, CP 193, I-73100,
  Lecce, Italy}
\affiliation{INFN, Sez. di Lecce, Via per Arnesano, CP 193, I-73100, Lecce, Italy}

\author{A. Read}
\affiliation{Department of Physics and Astronomy, Leicester University, Leicester LE1 7RH, U.K.}

\author{D. Licchelli}
\affiliation{Department of Mathematics and Physics {\it ``E. De Giorgi''}, University of Salento, Via per Arnesano, CP 193, I-73100,
  Lecce, Italy}

\author{G. Ingrosso}
\affiliation{Department of Mathematics and Physics {\it ``E. De Giorgi''}, University of Salento, Via per Arnesano, CP 193, I-73100,
  Lecce, Italy}

\affiliation{INFN, Sez. di Lecce, Via per Arnesano, CP 193, I-73100, Lecce, Italy}
\author{F. Convenga}
\affiliation{Department of Mathematics and Physics {\it ``E. De Giorgi''}, University of Salento, Via per Arnesano, CP 193, I-73100,
  Lecce, Italy}

\author{K. Boutsia}
\affiliation{Carnegie Observatories, Las Campanas Observatory, Colina El Pino, Casilla 601, La Serena, Chile}

\begin{abstract}
{SN 2015J was discovered on April 27th 2015 and is classified as a type IIn supernova. At first, it appeared 
to be an orphan SN candidate, i.e. without any clear identification of its host galaxy. Here, we present the analysis of the observations carried out 
{by the VLT 8-m class telescope with the FORS2 camera in the R band and the Magellan telescope (6.5 m) equipped with the IMACS Short-Camera (V and I filters) and 
the FourStar camera (Ks filter)}. We show that SN 2015J resides in what appears to be a 
very compact galaxy establishing a relation between the SN event and its natural host. We also present and discuss archival and new $X$-ray data 
centred on SN 2015J. At the time of the supernova explosion, Swift/XRT observations were made and a weak X-ray source was detected at the location of SN 2015J. Almost one year later, 
the same source was unambiguously identified during serendipitous observations by Swift/XRT and $XMM$-Newton, clearly showing an enhancement 
of the 0.3-10 keV band flux by a factor $\simeq 30$ with respect to the initial state. Swift/XRT observations 
show that the source is still active  in the $X$-rays at a level of $\simeq 0.05$ counts s$^{-1}$. The unabsorbed X-ray luminosity derived 
from the {\it XMM}-Newton slew and SWIFT observations, $L_{x}\simeq 5\times10^{41}$ erg s$^{-1}$, 
places SN 2015J among the brightest young supernovae in X-rays.}
\end{abstract}


\section{Introduction}

It is not obvious that supernova (SN) events may occur in a location of the sky without any associated host galaxy. As a matter of fact,
the Sternberg Astronomical Institute (SAI) supernova catalog \citep{Tsvetkov2004}, reports more than 5000 such events whose nature needs
to be addressed. In this respect, there is certainly the possibility that the host galaxy is not detected in the majority of the cases because of 
its low surface brightness or, in a more challenging hypothesis, the SN progenitors are hyper-velocity stars characterized by velocities as large as  $\simeq 700-1000$ km s$^{-1}$ 
(see e.g. \citealt{Martin2006}, \citealt{Brown2005,Brown2011,Brown2014}) which have escaped their host galaxy. In both cases, these supernovae are labeled as {\it orphan SN events}, unless
the host galaxy is identified in follow up observations.

SN 2015J is a supernova\footnote{The source is also labeled as SMTJ07350518-6907531.} which occurred on April 27th 2015 at the (J2000) coordinates RA = $07\hour:35\minute:05.18\second$ and DEC = $-69\degr:07\minute:53.1\arcsec$ and 
observed for the first time by \citet{Childress} with the 268-megapixel camera on the SkyMapper 1.3-m telescope 
at Siding Spring Observatory (Australia) as part of the SkyMapper Transient (SMT) survey \citep{Scalzo2017}. Source images acquired on April 2015 
(27.9 UT) had both $g$ and $r$ magnitudes of $19.3$. 

The SN light curve showed an initial peak of $r = 18.2$ on May (08.9 UT) before declining. The light curve then 
rose again up to a second peak at r = 16.8 on June 09.9 UT.  After a brief decline, it rose again and, as noted 
by the same authors (to whom we refer for more details), the last observation showed a magnitude of $r \simeq 16.0$ still rising.  
\citet{Childress} also acquired a 20 minute spectrum with the Wide Field Spectrograph (WiFeS, \citealt{Dopita}) which revealed a Type IIn SN\footnote{Type IIn supernovae are characterized by narrow emission 
Balmer lines in their spectra \citep{filippenko1997} and  show signatures of the
interaction between the SN ejecta and the
circumstellar medium (see, e.g., \citealt{Schlegel1990}). They form an heterogeneous object sample as their peak luminosity (which span more than two orders of magnitudes, \citealt{richardson2014}), 
depends on many factors such as the circumstellar medium density, the SN ejecta mass and input energy.} located
at redshift $z=0.0054$, corresponding to a distance of $\simeq 24.2$ Mpc (adopting the more recent cosmological parameters $H_0=67.15$ km s$^{-1}$ Mpc$^{-1}$, $\Omega_M=0.27$ and $\Omega_{\Lambda}=0.73$).

The target was also observed with the Australia Telescope Compact Array (ATCA) on June 2015 (29.1 UT) \citep{Ryder2015} revealing a radio 
source at a position consistent with that measured optically, and with fluxes of $0.07\pm0.02$ mJy and $0.10\pm0.03$ mJy at frequencies of $9.0$ and $5.5$ GHz, respectively. 

{When searching for an optical counterpart,  \citet{Childress} found no obvious host galaxy close to the supernovae location and noted that SN 2015J
might be associated with a group of galaxies at similar redshifts. In particular, NGC 2434 (at z=0.004637 as given by NED) is at a projected distance of $62$ kpc 
and NGC 2442 (z=0.004890 via NED) is at 166 kpc}. In addition, SN 2015J is close to
a $10^9$ M$_{\odot}$ gas cloud (HIPASS J0731-69, see \citealt{Ryder2001}). NGC 2442 seems to interact with
HIPASS J0731-69 and the whole complex can be considered as the result of a previous gravitational interaction. However, the two galaxies are too far 
to explain SN 2015J as originating from a hyper-velocity star progenitor that escaped one of them. It is then natural to expect that the SN occurred in a faint and small
galaxy not previously identified.

To address this issue, {we obtained $\simeq 0.98$ hour of net exposure time} on the VLT 8-m class telescope with the FORS2 camera in the R filter 
and a few exposures ({180 seconds each}) from the Magellan telescope (6.5 m) with the IMACS Short-Camera in the V and I filters and with the FourStar camera in the Ks band. The analysis of these observations, presented in this paper, 
allowed us to definitely show that SN 2015J resides in what appears to be a very compact galaxy, finally establishing a relation between the SN event and its natural host. 
After comparing the newly acquired data with archival ones (as the Digital Sky Survey -DSS- blue and infrared images
\footnote{The Digitized Sky Survey comprises a set of all-sky photographic surveys with the Palomar and UK Schmidt telescopes. The DSS data may be retrieved from \url{http://archive.eso.org/dss/dss}.} and the 
Catalina Sky Survey -CSS- data\footnote{The CSS, which uses a 1.5 meter telescope on the peak of Mt. Lemmon and a 68 cm telescope near Mt. Bigelow (both in the Tucson area, USA),
is a project to discover comets and asteroids and to search for near-Earth objects (NEOs). Further information on the CSS project are available 
at \url{http://www.lpl.arizona.edu/css/.}}) we realized that, prior to the SN explosion, the light of the SN progenitor contributed in a non negligible way to the host galaxy brightness {(see Section 2 for details)}.

Type IIn SN events are among the most luminous $X$-ray supernovae (see e.g. \citealt{chandra2012,chandra2015}), and indeed $X$-ray observations give important information about the SN itself and its environment.
The region of the sky around SN 2015J was observed many times since 2004 by the $XMM$-Newton satellite in slew mode but without any detection. Also the Chandra observatory did not detect any $X$-ray 
source during past observations. At the time of the supernova explosion, 
Swift/XRT observations were made and a weak source of X-rays was detected at the location of SN 2015J. Almost one year later, 
the same source was clearly identified during serendipitous observations by Swift/XRT and $XMM$-Newton, showing an enhancement 
of the 0.3-10 keV band flux by a factor $\simeq 30$ with respect to the initial state. We requested new Swift/XRT observations which, performed on March and July 2017, 
clearly show that the source is still active in the $X$-rays at a level of $\simeq 0.05$ counts s$^{-1}$, as will be discussed in Section 3.  

The paper is organized as follows: in Section \ref{sect:optical}, we report on archival (DSS and CSS) and newly acquired (VLT/FORS2 and Magellan/IMACS/FourStar) data, while leaving to Section \ref{sect:xray}
the analysis of all the available $X$-ray data. We finally present our conclusion in Section \ref{sect:conclusion}.

\section{SN 2015J: optical and near-IR data}
\label{sect:optical}
\begin{figure*}
\centering
\vspace{1cm}
{\includegraphics[width=2.0\columnwidth, ]{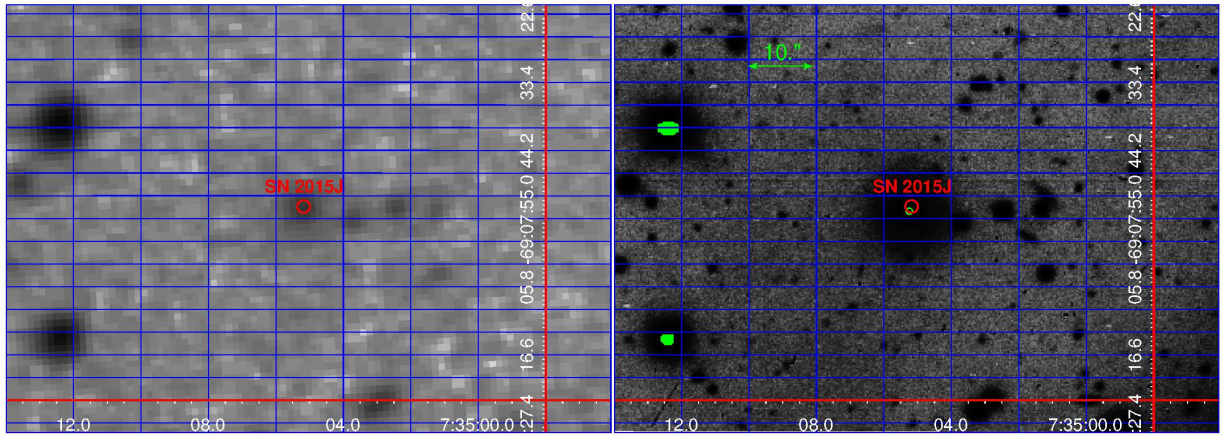}}
\caption{The DSS $F_{ph}$ band image (left panel) and the image acquired by the VLT/FORS2 in R band (right panel) of the region toward SN 2015J are shown. 
The red circle is centred on SN 2015J. The green 
circle (with a $0.5\arcsec$ radius and centred on the centroid of the corresponding brightness surface) appearing on the right panel corresponds to the source detected 
in the the VLT/FORS2 image.}
\label{figure1}
\end{figure*}

The absence of a clear identification of the SN 2015J host galaxy makes the event a possible orphan SN candidate. With the aim to find if any extended optical 
counterpart does exist, we searched for past DSS images acquired in the direction of the target. At the coordinates of the supernova explosion, a source was clearly 
identified in the DSS survey in the $J_{ph}$ ($340-590$ nm), $F_{ph}$ ($590-715$ nm), and $N_{ph}$ ($700-970$ nm) images. 
In Figure \ref{figure1} (left panel), we show the image in 
the $F_{ph}$ band of the DSS field towards the SN 2015J. The red circle (with radius of $\simeq$ 1$\arcsec$) is centred on the nominal coordinates of SN 2015J. 

The {\it Catalogs and Surveys Branch} of the Space Telescope Science Institute digitized the photographic plates from the DSS to produce the Guide Star Catalog
which ultimately contains positions, proper motions, classifications, and magnitudes in multiple bands for almost a billion objects down to approximately 
$J_{ph}=21$, and $F_{ph}=20$. The source identified at the SN position has magnitudes in the photographic 
bands $F_{ph}$, $J_{ph}$  and $N_{ph}$  of 18.29, 19.61 and 18.26, respectively. 

{The source identified in the DSS images appears to be point-like and very faint thus opening two possibilities: the DSS source is the unresolved SN 2015J 
host galaxy or, in the orphan supernovae scenario \citep{zinn}, it is the progenitor of SN 2015J which is believed to be an Eta-Carinae like star 
(see, e.g. \citealt{galyam}) which, having an absolute magnitude of $\simeq -12$, at the distance of $\simeq 24.2$ Mpc (as that of the observed supernova), 
would have a visual magnitude of about 19.5, not far from that corresponding to the DSS source. 
}

{To distinguish between these two possibilities}, we requested observing time at the VLT telescope equipped with the FORS2 camera in the R filter (550 nm - 800 nm) and at the Magellan telescope with the IMACS (V and I filters) and 
FourStar (Ks filter). As far as the VLT/FORS2 data is concerned, each single frame (acquired in December 2016) was corrected using standard procedures (bias, dark current and flat field corrections) and then geometrically aligned in order to get a 
calibrated and clean image (in counts s$^{-1}$) as an averaged sum. The VLT/FORS2 R band image resulted in a total exposure time of $0.98$ hours. 
In Figure \ref{figure1} (right panel), we give a zoom (with histogram equalization) around the target.
The green  circles (each with a $0.5\arcsec$ radius and centred on the centroid of the corresponding brightness surface) on the right panel correspond to some of the 
sources detected in the the VLT/FORS2 image by using the SExtractor code. We extracted the aperture photometry with a source extraction radius of $\simeq 4\arcsec$ 
(i.e. well above the FWHM of the FORS2 camera) while the background was evaluated locally. After calibrating the photometry with the GSC2.3 catalogue 
(although the photographic band of the DSS plates is similar (but not exactly the same) to the FORS2 R band), the target
source resulted to have magnitude (Vega system) in the R band of $18.80\pm 0.20$. 
\begin{figure*}
\centering
\vspace{2cm}
{\includegraphics[width=1.6\columnwidth, ]{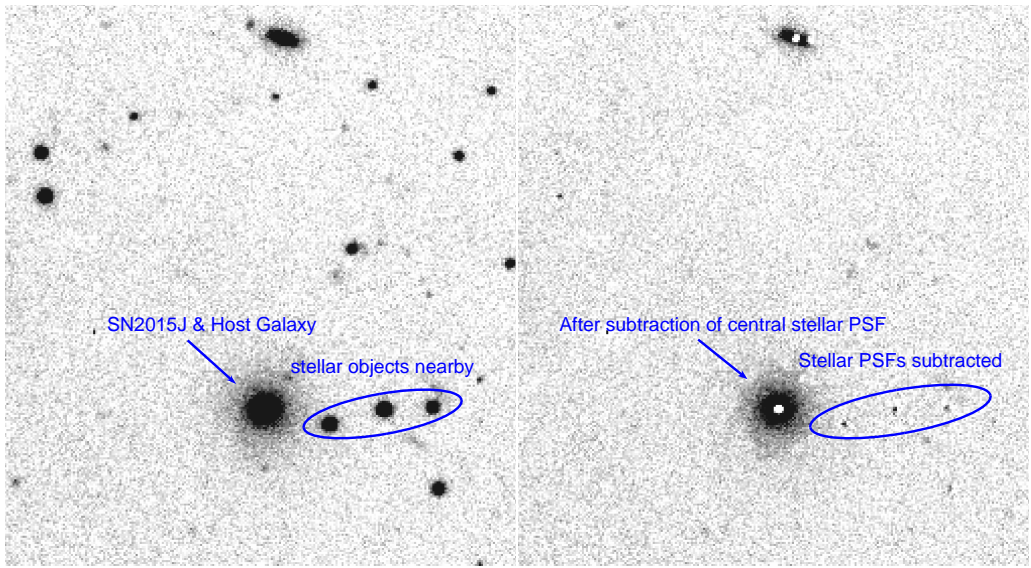}}
\caption{The VLT/FORS2 image of the region around SN 2015J before (left panel) and after (right panel) subtracting the PSF obtained by using DAOPHOT. 
As it is clear, the target source is an extended object, probaly a dwarf galaxy (see text for details).}
\label{figuresubtractions}
\end{figure*}
{The target appeared in the VLT image as a clearly extended source, and we verified its extension 
by deriving the point spread function (PSF) of the frame 
by using the DAOPHOT package (\citealt{stetson}). In particular, we invoked the {\it find} and {\it photometry} routines 
in order to determine the position and aperture photometry of any source above a certain background threshold. We then selected an adequate number of 
round and isolated (likely PSF) stars. Hence, the DAOPHOT {\it psf} routine allowed us to estimate an image PSF characterized by a FWHM of $\simeq 0.65\arcsec$. 
After subtracting the PSF from the image, we were left with an image of residuals (Figure \ref{figuresubtractions}) 
in which all the star-like objects are removed. Note that large deviations (from what expected 
for a point-like object) appear at the location of the source and this is convincing evidence we are dealing with an extended object\footnote{Note that a diffuse brightness surrounds the target and, based on the K 
image of Figure \ref{figure2}, an emission peak
(the {\it knot} with magnitude $18.77 \pm 0.18 $) is also present in a position located at the South-East from the core.}.

As a matter of fact the SExtractor code classified the target as a galaxy (with ellipticity parameter $0.127$ and S/G classifier output of $0.03$). 
The extraction radius used in order to encompass the visible halo of the galaxy is $\simeq 4\arcsec$, corresponding 
to a linear length of $\simeq 0.5$ kpc for an estimated distance of $24.2$ Mpc as that of SN 2015J.

{The distance between SN 2015J and the centroid coordinates of the closest source 
detected in the FORS2 image is $\simeq 0.8\arcsec$. 

}

{The IMACS and FourStar images were reduced following standard procedures which allowed us to estimate the V, I and Ks magnitude (Vega system) {of the host galaxy (possibly with a contribution of the SN, 
see next discussion)} to be $19.68\pm 0.07$, 
$18.61\pm 0.08$, and $15.80 \pm 0.15$, respectively. Note that the photometry of the K band image was obtained by calibrating on the 2MASS catalogue sources found 
within the field of view, while V and I magnitudes have been calibrated using the standard stars in Landolt field RU149 observed in the same night.} 
As an example, Figure \ref{figure2} shows a zoom of the field of view around the target as appearing in the FourStar Ks band. 

{From the discussion above, the presence of the SN host galaxy is evident thus supporting the scenario of SN 2015J being a normal 
SN occurred in a compact galaxy at a distance of about $24.2$ Mpc. {As a last remark on the host galaxy, in
the NASA's Wide-field Infrared Survey Explorer (WISE, \citealt{wright2010}) all sky data, a source was detected in 2010, i.e. well before the supernova explosion. The source had 
W1 (3.4 $\mu$m) and W2 (4.6 $\mu$m) magnitudes of $\simeq 15.91$ and $\simeq 15.89$, respectively.}}

\begin{figure*}
\centering
\vspace{-1cm}
{\includegraphics[width=1.3\columnwidth, ]{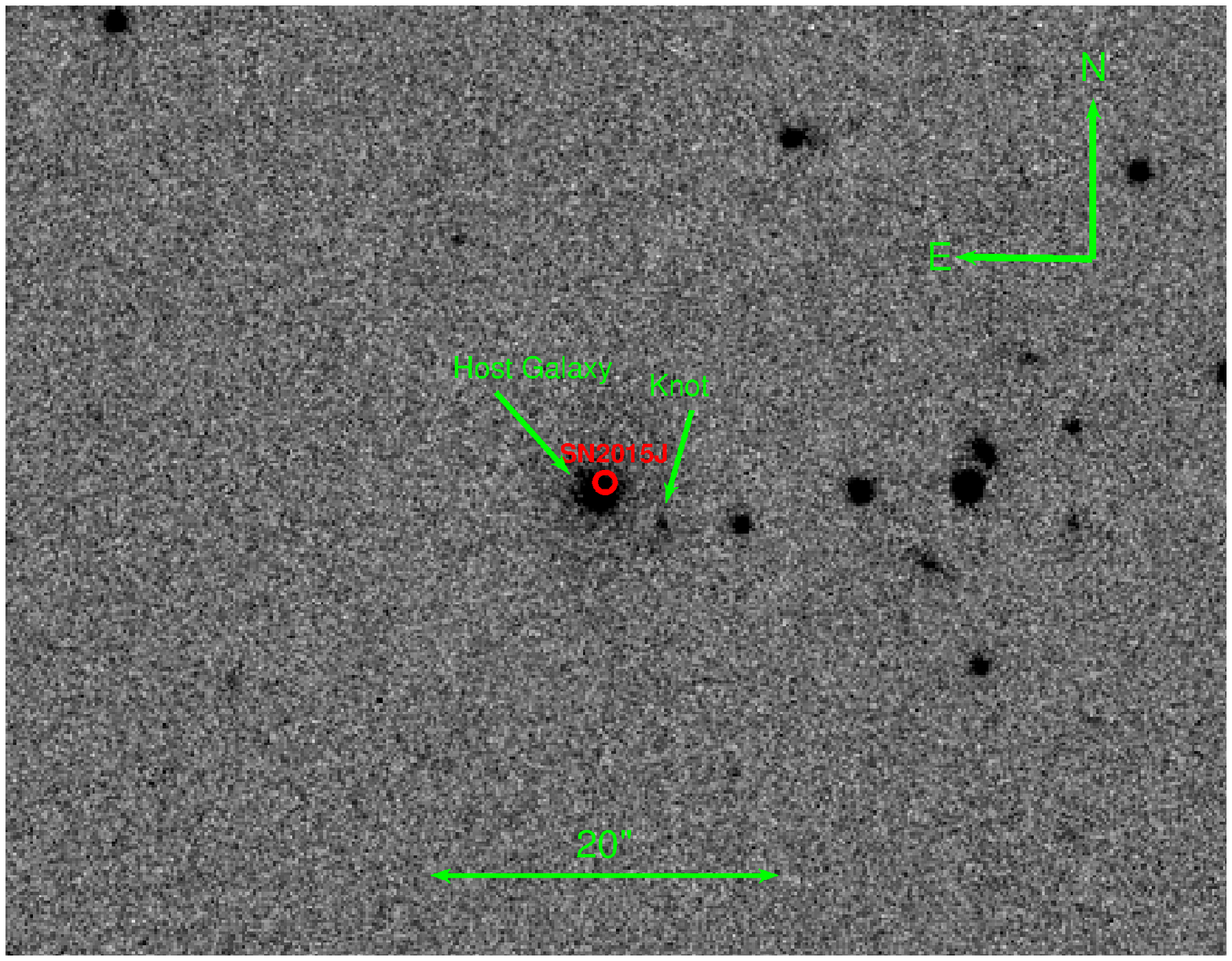}}
\caption{The FourStar Ks image of the galaxy host of SN 2015J.}
\label{figure2}
\end{figure*}
\begin{figure*}
\centering
{\includegraphics[width=0.8\textwidth, ]{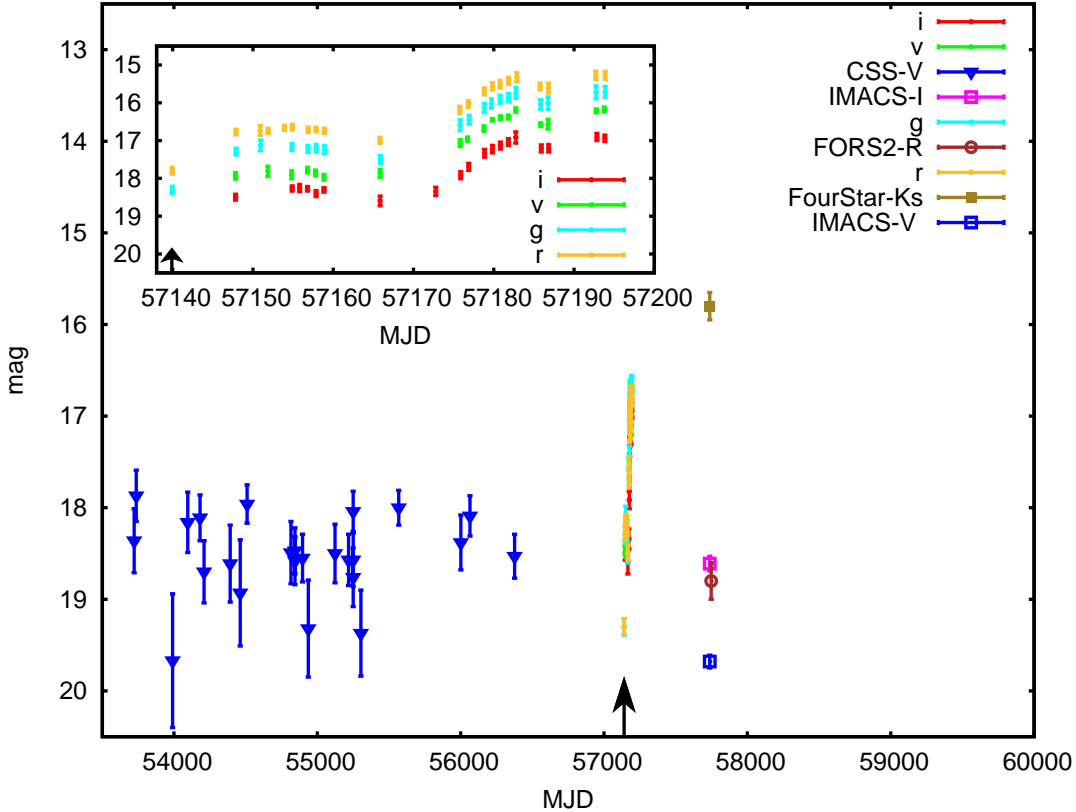}}
\caption{The multi-band light curve of the target including the CSS data (filled triangles, see text for details) is shown. The inset shows a zoom of the light curve (dots as taken directly 
from \citealt{Childress}) about the time of the SN event which occurred 
on April 2015 27.9 UT (corresponding to MJD=57139.9, as also indicated by the black arrow). From $i$ to $r$ band we applied a recursively offset of $0.5$ magnitude for each light curve shown in the inset.}
\label{figurehist}
\end{figure*}
The multi-band light curve of the target including the CSS data, the data acquired during the SN event {(as extracted from the figures in \citealt{Childress})}, 
and the data obtained from VLT and MAGELLAN, is shown in 
Fig. \ref{figurehist}. The inset shows a zoom of the light curve about the time of the SN event {with the black arrow indicating the time of the SN explosion}. 
{The comparison between the V band magnitude as estimated from the Magellan IMACS data {($\simeq 19.68\pm 0.07$) and the mean value
observed in the CSS data before the SN explosion ($\simeq 18.52\pm 0.45$)} allow us to conclude that the progenitor of the SN event contributed in a non-negligible way to the host galaxy brightness. In particular, 
we find ${\rm F_{PSN}/F_{G}}\simeq 1.9$, being ${\rm F_{PSN}}$ and ${\rm F_{G}}$ the fluxes of the pre SN event and host galaxy, respectively.}

\section{SN 2015J: the $X$-ray view}
\label{sect:xray}

The location of SN 2015J was observed in several XMM-Newton slews since 2004 but no source was detected. A Chandra observation 
(OBSid 2923 antecedent to the SN) did not detect any X-ray source at the position of SN 2015J and put an upper limit to the quiescent X-ray flux 
of $F_{0.2-8 {\rm keV}}\simeq 6\times 10^{-15}$ erg cm$^{-2}$ s$^{-1}$, about 100 times deeper
than the XMM-Newton slew upper limits. 

{SN 2015J  was observed by Swift/XRT in several occasions soon after the SN explosion (see Table \ref{tablelog}). The
Swift data have been analyzed by using the standard procedures described in \citet{burrows} with the
latest calibration files. We processed the XRT data with the {\it XRT-Pipeline} (v.0.12.6) task and we applied standard 
screening criteria by using the FTOOL (Heasoft v. 6.19). The source spectra have been
extracted from a circular region centred on the target nominal coordinates (with radius of 40 arcseconds) by using the {\it xselect}
routine, while the background spectra were obtained from an annulus with external radius of 60 arcseconds. We then used the {\it xrtmkarf} script to create 
the ancillary response files and imported the grouped spectra within {\it XSPEC} \citep{arnaud} for the spectral analysis.

As shown in Table \ref{tablelog}, the Swift/XRT count rate of the target source was $\simeq 7.0\times 10^{-3}$ counts s$^{-1}$ 
in the $0.3-10$ keV energy band during the SN explosion (26th August 2015, 07.46 UT). 

The source was serendipitously observed about one year after the SN explosion (in April 2016) by Swift/XRT which detected the source with an increased 
count rate of $\simeq 5.0 \times 10^{-2}$ 
counts s$^{-1}$. On 15th September 2016, observing for 7.2 seconds with the EPIC-pn camera ($XMM$-Newton observation in slew mode, XMMSL1 J073504.6-690752, 
OBSid $9307100003$), a count rate of $1.7 \pm 0.4$ counts s$^{-1}$ was observed from the source.

 \renewcommand{\thefootnote}{\fnsymbol{footnote}}
 \renewcommand{\arraystretch}{0.81}

 \begin{longtable*}{|c|c|c|c|}
 \caption{The log of the Swift/XRT and {\it XMM}-Newton$^{*}$ (11th row) observations. We give the observation identification number, the (mid) time of exsposure, the live time of each observation 
 and the source count rate in the $0.3-10$ keV band. }
 \label{tablelog}\\
 \hline\hline 
    \multicolumn{1}{c}{\textbf{OBS ID}} &
     \multicolumn{1}{c}{\textbf{T}} &
    \multicolumn{1}{c}{\textbf{Live Time}} &
    \multicolumn{1}{c}{\textbf{Rate}} \\
    \multicolumn{1}{c}{\textbf{ }} &
    \multicolumn{1}{c}{\textbf{MJD}} &
    \multicolumn{1}{c}{\textbf{s}} &
    \multicolumn{1}{c}{\textbf{count s$^{-1}$}} \\
    \hline\hline
 \endfirsthead
 \multicolumn{4}{c}{{\tablename} \thetable{} -- Continued}\\
 \hline\hline
     \multicolumn{1}{c}{\textbf{OBS ID}} &
     \multicolumn{1}{c}{\textbf{T}} &
    \multicolumn{1}{c}{\textbf{Live Time}} &
    \multicolumn{1}{c}{\textbf{Rate}} \\
    \multicolumn{1}{c}{\textbf{ }} &
    \multicolumn{1}{c}{\textbf{MJD}} &
    \multicolumn{1}{c}{\textbf{s}} &
    \multicolumn{1}{c}{\textbf{counts s$^{-1}$}} \\
 \hline\hline 
 \endhead
 \multicolumn{4}{c}{{Continued on Next Page\ldots}} \\
 \endfoot
 \hline\hline
 \endlastfoot

$00033857002$  &   $57201.532$   &    $1743.12$ & $ (2.7\pm 1.5)\times 10^{-3}$\\
$00033857003$  &   $57204.789$   &    $1945.40$ & $ (5.3\pm 1.8)\times 10^{-3}$\\
$00033857004$  &   $57207.045$   &    $2349.50$ & $ (4.6\pm 1.6)\times 10^{-3}$\\
$00033857005$  &   $57210.735$   &    $1860.50$ & $ (4.6\pm 1.8)\times 10^{-3}$\\
$00033857006$  &   $57221.539$   &    $1640.72$ & $ (3.6\pm 1.6)\times 10^{-3}$\\
$00033857008$  &   $57236.589$   &    $1937.91$ & $ (3.9\pm 1.7)\times 10^{-3}$\\
$00033857009$  &   $57244.098$   &    $1997.83$ & $ (5.9\pm 1.9)\times 10^{-3}$\\
$00033857010$  &   $57252.550$   &    $1965.38$ & $ (6.7\pm 2.0)\times 10^{-3}$\\
$00033857011$  &   $57260.324$   &    $2017.83$ & $ (7.0\pm 2.0)\times 10^{-3}$\\
$07002410001$  &   $57499.501$   &    $59.94  $ & $ (5.0\pm 2.9)\times 10^{-2}$\\
$9307100003^{*}$  &   $57646.527$   &    $7.2    $ & $ (9.0\pm 3.0)\times 10^{-2}$\\
$00033857012$  &   $57820.517$   &    $1853.00$ & $ (4.9\pm 0.5)\times 10^{-2}$\\
$00033857013$  &   $57946.721$   &    $1313.58$ & $ (5.4\pm 0.6)\times 10^{-2}$
\end{longtable*}
 \normalsize
 \renewcommand{\thefootnote}{\arabic{footnote}}
\renewcommand{\arraystretch}{1.0}

{The spectrum was soft\footnote{Note however that most of the Type IIn supernovae have typically hard $X$-ray spectra (see, e.g., \citealt{chandra2012a,chandra2012,Stritzinger2012}).}, i.e. characterized by an equivalent power-law slope $\Gamma \simeq 3$ or black-body $kT\simeq 0.13$ keV and absorbed flux of $F_{0.3-2 {\rm keV}}= (2.1^{+1.1}_{-1.0})\times 10^{-12}$
erg cm$^{-2}$ s$^{-1}$. The unabsorbed flux, assuming a column density of $2\times 10^{21}$ cm$^{-2}$ (\citealt{willingale2013}), 
is $F_{0.3-2 {\rm keV}} = (7.5^{+3.9}_{-3.5})\times 10^{-12}$ erg cm$^{-2}$ s$^{-1}$  corresponding to a luminosity of 
$L_{0.3-2 {\rm keV}}= (5.2^{+2.7}_{2.5})\times 10^{41}$ erg s$^{-1}$ for the estimated distance ($\simeq 24.2$ Mpc) of the host galaxy.
}
\begin{figure*}
\centering
{\includegraphics[width=1.2\columnwidth, ]{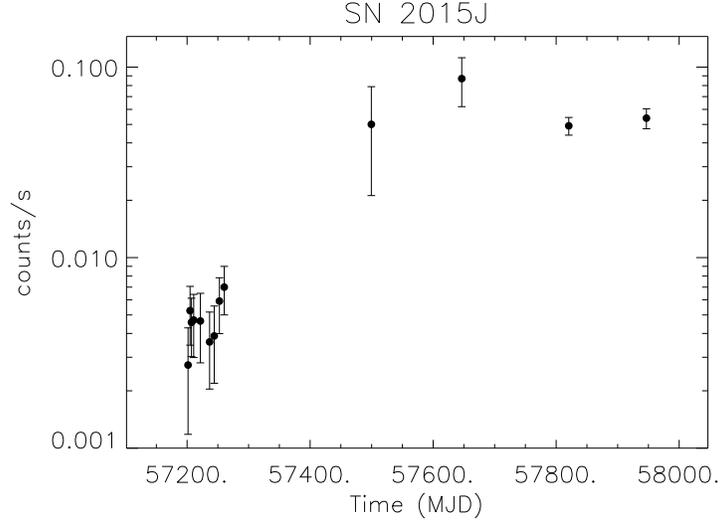}}
\caption{The $0.3-10$ keV light curve of SN 2015J. The last data point (acquired by SWIFT/XRT in July 2017) shows that the SN event is still active in the $X$-rays after more than one
 year at a level of $\simeq 0.05$ counts s$^{-1}$. The XMM slew flux at MJD=57646.5 has been converted into a Swift count rate (see text for details).
 }
\label{figure3}
\end{figure*}

{
In order to determine the behaviour of the X-ray light curve, we requested new observations of the source by Swift/XRT 
in March 2017 (Obs ID $00033857012$) and again in July 2017 (Obs ID $00033857013$) 
and verified that the source is still active in the X-rays at a level of  
$\simeq 5.4\times 10^{-2}$ counts s$^{-1}$. In Figure \ref{figure3}, we show the 0.3-10 keV light curve 
in terms of the Swift/XRT PC count rate, where the $XMM$-Newton slew point has been converted to the Swift count rate using the WebPIMMS tool\footnote{WebPIMMS is available at 
url{https://heasarc.gsfc.nasa.gov/cgi-bin/Tools/w3pimms/w3pimms.pl}} and a power-law spectral model (with $\Gamma=3$). 
The X-ray flux has been consistently high between August 2016 and the latest observation in July 2017. 
}


{
A spectrum was extracted from the March 2017 Swift XRT observation and rebinned in order to have at minimum 5 counts per channel with {\it grppha}. 
The resulting spectrum (presented in Figure \ref{figurespectrumXa}) 
appears to be soft. In fact, assuming an aborbed power law
with $nH=2\times 10^{21}$ cm$^{-2}$, one gets a power slope of $\Gamma=4.0 \pm 0.5$ ($\chi^2=0.6$ for 15 d.o.f.). The absorbed flux in the 0.3-2 keV band is 
$(1.1\pm 0.2)\times 10^{-12}$ ergs cm$^{-2}$ s$^{-1}$ corresponding to an unabsorbed flux of $(6.2\pm1.1)\times 10^{-12}$ 
ergs cm$^{-2}$ s$^{-1}$. The associated luminosity is $L_{0.3-2 {\rm keV}}= (4.3 \pm 0.8) \times 10^{41}$ erg s$^{-1}$.
}

{
As far as the last Swift XRT observation (July 2017), the spectrum (see Figure \ref{figurespectrumXb}) can be described by 
an absorbed power law with $nH=2\times 10^{21}$ cm$^{-2}$ and power slope of $\Gamma=3.5 \pm 0.5$ ($\chi^2=0.7$ for 11 d.o.f.). The absorbed flux in the 0.3-2 keV band is 
$(1.2\pm 0.3)\times 10^{-12}$ ergs cm$^{-2}$ s$^{-1}$ corresponding to an unabsorbed flux of $(5.2\pm1.3)\times 10^{-12}$ 
ergs cm$^{-2}$ s$^{-1}$. The associated luminosity is $L_{0.3-2 {\rm keV}}= (3.6 \pm 0.9) \times 10^{41}$ erg s$^{-1}$, thus implying an approximately constant $X$-ray luminosity.
}
}

For completeness, the SWIFT/UVOT images (in the U filter centred at 346.5 nm) was analyzed using the {\it uvotdetect} and {\it uvotsource} scripts. By performing aperture photometry
with a radius of 5$\arcsec$, we found a U magnitude of $19.1\pm 0.1$.  
\begin{figure*}
\centering
 \vspace{+1.5cm}
{\includegraphics[width=1.1\columnwidth, angle=-90, origin=c ]{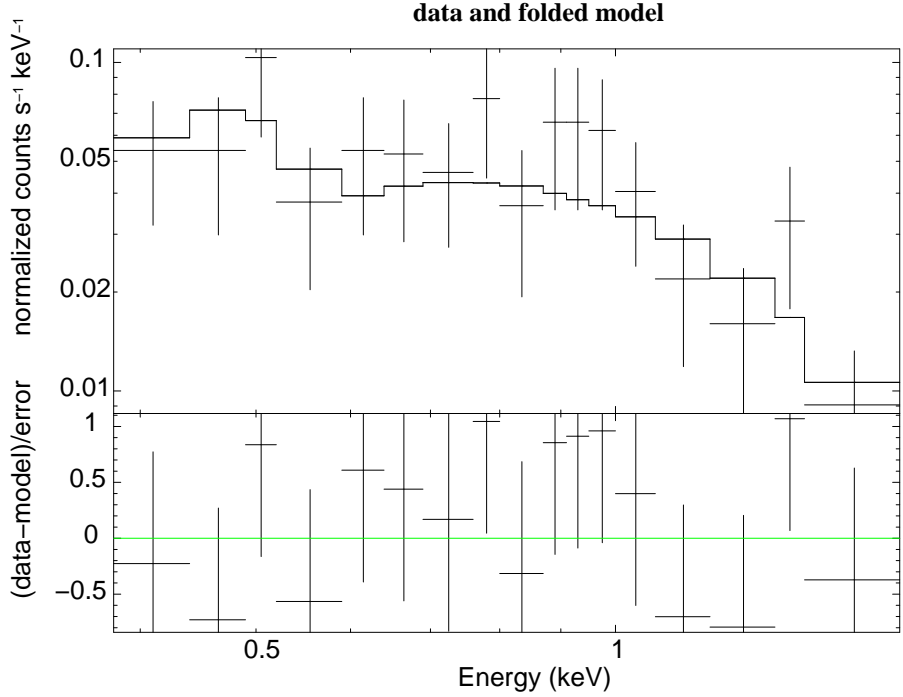}}
\caption{The 0.3-1.7 keV spectrum of SN 2015J as observed by Swift/XRT (OBSid 00033857012).}
\label{figurespectrumXa}
\end{figure*}
\begin{figure*}
\centering
 \vspace{+1.5cm}
{\includegraphics[width=1.1\columnwidth, angle=-90, origin=c ]{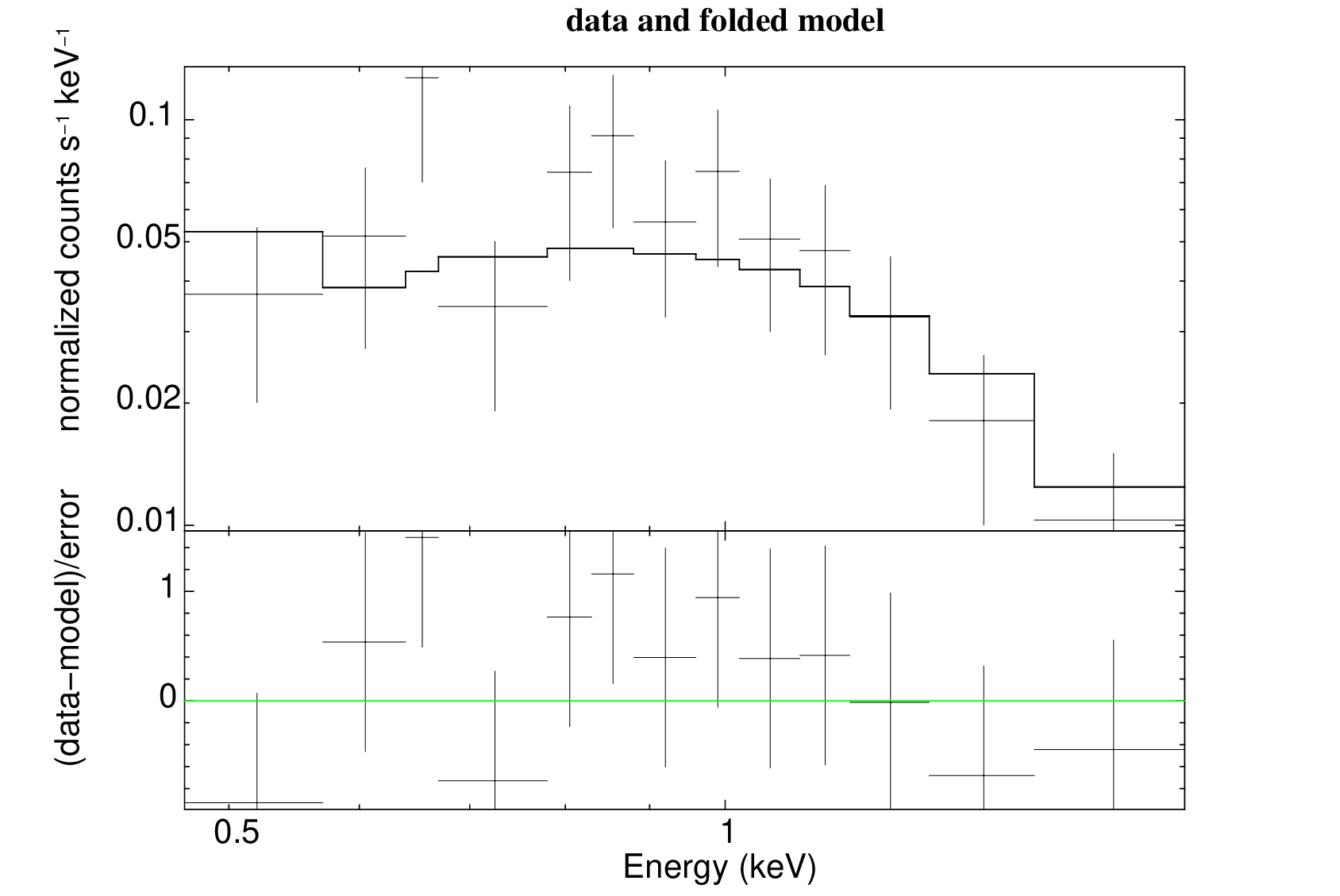}}
\caption{The 0.3-1.7 keV spectrum of SN 2015J as observed by Swift/XRT (OBSid 00033857013).}
\label{figurespectrumXb}
\end{figure*}

\section{Results and discussion}
\label{sect:conclusion}

We presented the analysis of the observations acquired by the VLT (R band) and Magellan (V, I, and K bands) telescopes as well as the $X$-ray data 
from the {\it XMM}-Newton (slew mode) and Swift satellites towards the SN 2015J location. {Optical and near-IR observations allowed us to discover of the SN host galaxy 
which appears to be a compact dwarf galaxy with a size of $\simeq 1$ kpc at a distance of $24.2$ Mpc. 

The $X$-ray data show that the SN is still active after about two years after 
the explosion and the the derived unabsorbed X-ray luminosity  $L_{0.3-2 {\rm keV}}\simeq 5.2\times10^{41}$ erg s$^{-1}$, 
places SN 2015J among the most luminous young SNe (see \citealt{Dwarkadas2011} for a list of SNe observed in X-rays).  We note, for comparison, that 
SN 2010jl was classified as a type IIn supernova with unabsorbed 0.2—10 keV luminosity of about $7\times10^{41}$ erg s$^{-1}$ at 2 and 12 
months after the event \citep{chandra2012a}. Its spectra could be described by a MEKAL model characterized by a high temperature, $kT > 8$ keV, absorbed by a column which dropped 
from $nH \simeq 10^{24}$ to $nH\simeq 3\times10^{23}$ cm$^{-2}$ between the observations. Also SN2006jd showed a large X-ray luminosity (about $4\times10^{41}$ erg s$^{-1}$) 
for several years after the explosion \citep{Stritzinger2012}. The temperature was also very high here, $kT>20$ keV with an intrinsic column of $nH\simeq 10^{21}$ cm$^{-2}$ 
\citep{chandra2012}. In these type IIn SNe, the high luminosity has been attributed to a forward shock, from ejecta expanding at several 
thousands km s$^{-1}$ into a dense circumstellar medium expelled by the progenitor star in previous shell ejection episodes.

In this respect, we note that SN 2015J has properties in common with these two SNe  as regards the optical spectrum and the X-ray luminosity of a 
few $\times10^{41}$ ergs/s which persisted for more than a year. However, its X-ray spectrum, as deduced from the admittedly 
low statistics of the Swift observations, appears to be much softer than the other type IIn SNe. In fact, when fitting the last SWIFT/XRT data (July 2017) with an 
absorbed MEKAL model (i.e., {\it wabs*zwabs*mekal} within XSPEC, with the {\it wabs} component 
accounting for the galactic hydrogen column density and {\it zwabs} for any intrinsic absorption) the best-temperature is $kT=0.17^{+0.4}_{-0.04}$ keV 
($1\sigma$ errors) with an upper limit of $\simeq 1.1$ keV with a $3-\sigma$ confidence level (see also Fig. \ref{figurecontour}). Note however that the intrinsic column 
density remains unconstrained (with maximum allowed value of $\simeq 1.2\times 10^{22}$ cm$^{-2}$).
\begin{figure*}
\centering
\vspace{+1.5cm}
{\includegraphics[width=1.0\columnwidth,angle=-90, origin=c ]{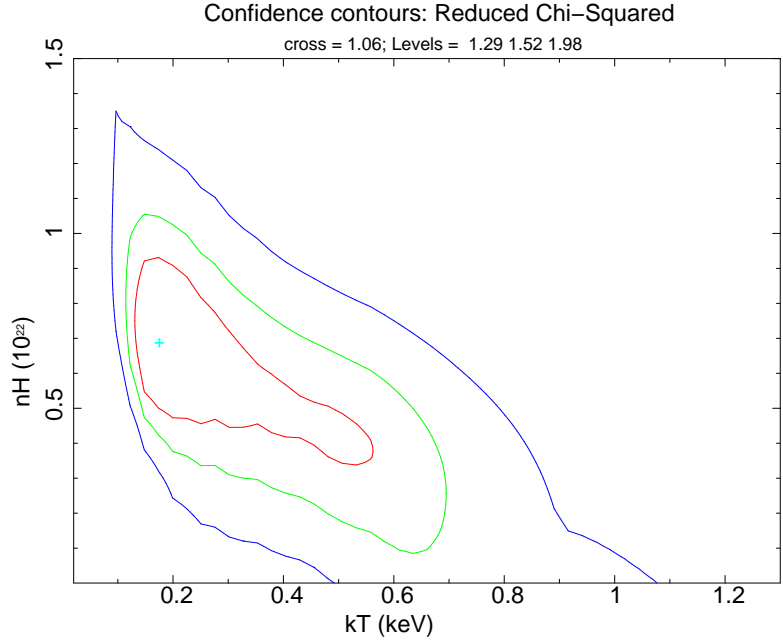}}
\caption{A contour plot of the temperature and absorption column parameters from a 
fit of an intrinsically-absorbed MEKAL model to the Swift observation of July 2017 
(assuming $z=0.0054$ and galactic column density $nH=2.0\times 10^{21}$ cm$^{-2}$). Contours mark the 1,2 and 3-sigma errors.
}
\label{figurecontour}
\end{figure*}

A detailed discussion about the evolution of the $X$-ray light curve from young SNe can be found in \citet{Dwarkadas2010} to whom we refer for more details. 
Here we comment that 
the expansion of a SN shock wave into the ambient medium produces forward and reverse shocks that heat up the gas to sufficiently large temperatures and produce $X$-rays. The expected $X$-ray luminosity is then 
\begin{equation}
L_X\sim n_e^2 \Lambda V,
\end{equation}
where $n_e$ is the electron number density, $\Lambda$ the cooling function and $V$ the volume of gas involved in the emission. Assuming that the surrounding medium number density goes as $r^{-s}$ (with $s=2$ in the case of 
a steady wind), that the emission originates from a thin shell with size $\Delta r \propto r$ (as in the self-similar case) at distance $r$ from the SN and that $\Lambda\propto r/t$, one gets
\begin{equation}
L_X\sim \frac{r^{4-2s}}{t}.
\end{equation}
Therefore, the resulting $X$-ray emission would scale as $t^{-1}$ in the steady wind scenario. Following this discussion, we fitted the last three data points of the $0.3-2$ keV light curve 
assuming a model of the form $L_x=at^b$, where $a$ and $b$ are considered free fit parameters. Fitting the log-log data with the previous relation gave a best fit power-law index $b=-0.98\pm1.6$. 
Although, due to the large associated uncertainty, the light curve is consistent with a flat slope, 
the central value of the power-law index seems to be reminiscent of a steady wind whose density decreases as $r^{-2}$.

However, the column density estimated above is far below the value needed to create the X-ray emission. It is possible that we are observing the shocked emission through a 
hole in the local gas, as suggested in the case of SN 2006jd by \citet{chandra2012} (see also \citealt{katsuda}). In addition, the soft spectrum is difficult to be reconciled 
with the high observed $X$-ray luminosity if it is generated by shocked material, thus opening to other possible interpretations.

For example, a possibility could be that some part of the material ejected might have remained bound to the compact remnant and fallen back at later times (see, e.g. \citealt{dexter2013}).

Alternatively, a tidal disruption event
might mimic a SN explosion. For example, the superluminous SN event
ASASSN-15lh \citep{Dong2016} has been reinterpreted
as the tidal disruption of a star by a
rapidly spinning $10^8$ M$_{\odot}$ black hole by \citet{Leloudas2016} because of the temperature evolution,
the total luminosity of the event and the metal-rich host galaxy.
Roughly constant, soft ($\Gamma=3$ or $kT=0.17$ keV),
X-ray emission has been seen from the position
of ASSASN-15lh with a $L_X\simeq 2-8\times 10^{41}$ erg s$^{-1}$
lasting for several hundred days \citep{Margutti2017}.

The power-law index b of the post-peak SN 2015J X-ray light
curve is flatter, but still consistent with the canonical TDE decay curve of  $t^{-5/3}$,
although we note that more slowly evolving TDE light curves are predicted to be common \citep{Guillochon2013,Guillochon2015}.

Therefore, future photometric and  high precision spectroscopic observations in $X$-rays are important 
for addressing the issues that still remain open on SN 2015J since, if confirmed, it is one of the brightest type II SNe ever observed.
}

\section*{Acknowledgements}
We acknowledge the support by the INFN projects TAsP (Theoretical Astroparticle Physics Project) and EUCLID. 
We warmly thank the anonymous Referee for his/her suggestions that improved the paper.

\end{document}